\documentclass[letterpaper]{aa}
\usepackage{graphics}
\usepackage{txfonts}
\usepackage{natbib}

\begin{document}

\title
{Variable iron-line emission near the black hole of
Markarian\,766
\thanks{
Based on observations obtained with {{\it XMM-Newton}}, 
an ESA science mission funded by ESA Member States and NASA. TJT acknowledges
support from NASA grant NNG05GL03G and MD support from  
Czech Science Foundation grant GACR 205/05/P525.
}
}
\subtitle{}

\titlerunning{Iron-line emission from Mrk\,766}

\author
{L.\ Miller\inst{1} \and 
T.\ J.\ Turner\inst{2,3} \and 
J.\ N.\ Reeves\inst{3,4} \and 
I.\ M.\ George\inst{2,3} \and
D.\ Porquet\inst{5} \and
K.\ Nandra\inst{6} \and
M.\ Dovciak\inst{7}
}

\authorrunning{L.\ Miller et al.\ }

\institute{Dept. of Physics, University of Oxford, 
Denys Wilkinson Building, Keble Road, Oxford OX1 3RH, U.K.
\and 
Dept. of Physics, University 
of Maryland Baltimore County, 1000 Hilltop Circle, Baltimore, MD 21250
\and 
Code 662,  Exploration of the Universe Division,   
NASA/GSFC, Greenbelt, MD 20771
\and
Dept. of Physics and Astronomy, Johns Hopkins University, 3400 N 
Charles Street, Baltimore, MD 21218
\and
Max-Planck-Institut f\"{u}r extraterrestrische Physik, Postfach 1312, 85741, Garching, Germany
\and
Astrophysics Group, Imperial College London, Blackett Laboratory, Prince Consort Road, London SW7 2AW, U.K.
\and
Astronomical Institute, Academy of Sciences, Bocn\'{i} II, 141 31 Prague, Czech Republic
}

\date{Received / Accepted}

\abstract
{}
{We investigate the link between ionised Fe X-ray line emission and continuum
emission in a bright nearby AGN, Mrk\,766.
}  
{A new long ($433$\,ks) {{\em XMM-Newton}} observation is analysed, together with
archival data from 2000 and 2001.  The contribution from ionised line emission
is measured and its time variations on short (5--20\,ks) timescales are correlated
with the continuum emission.
}
{The ionised line flux is found to be highly variable and to be strongly correlated
with the continuum flux, demonstrating an origin for the ionised line emission that is
co-located with the continuum emission. Most likely the emission is ionised reflection 
from the accretion disc within a few A.U. of the central black hole, and its detection
marks the first time that such an origin has been identified other than by fitting to 
spectral line profiles. Future observations may be able to measure a time
lag and hence achieve reverberation mapping of AGN at X-ray energies.
}
{}

\keywords{accretion; galaxies: active}

\maketitle

\section{Introduction}

Active galaxies are thought to be powered by accretion onto
a $10^{6-9}$\,M$_{\odot}$ black hole, forming an accretion disc whose 
hot inner region produces UV radiation.  X-rays are most likely
produced by inverse Compton scattering 
\citep{hm93}.  A significant fraction of the X-rays
shines onto the disk, producing a 
``reflection spectrum'', whose main observable feature below 10\,keV is
Fe\,K$\alpha$ emission emitted by fluorescence or recombination processes at
6.4-7\,keV \citep{gf91,pounds90}. Near the black hole, gravitational redshift may lead
to a broad and asymmetric emission-line profile \citep{fab89} as first observed in
active galaxies with ASCA \citep{tan95,nandra97}.  
More detailed studies (e.g. \citealt{wilms}, \citealt{reynolds})
have since been 
made with XMM-Newton (\citealt{jansen}, hereafter {\it XMM})
but have produced a puzzling result:
despite the continuum X-ray source being highly time-variable, the
asymmetric emission thought to be the redshifted reflected iron-line
is not \citep{min03,vaughanfabian}.  
In some active galaxies the non-redshifted component of
Fe\,K$\alpha$ emission has been found to be variable 
(e.g. \citealt{nandra00}, \citealt{vaughan01}), but there has 
so far been no strong evidence
for the tight correlation between emission-line and continuum
variations on short timescales 
that should be expected if the emitting regions are
spatially co-located.
 
Analysis of a 2001 {\it XMM} observation
of Mrk~766, a narrow-line Seyfert 1 galaxy with $z=0.0129$
\citep{osterbrock},
provided evidence for Fe emission shifting in energy on
timescales of tens \citep{t04} to a hundred ks (\citealt{turner06}, hereafter T06), 
which if interpreted as being
Doppler shifts from orbiting material would place the emission 
$\sim 100 r_g$ ($r_g=GM/c^2$) from the central black
hole (T06). 
In 2005 a total of 433\,ks of new data were obtained with {\it XMM}, and
we present here one of the first results,
the detection of variable ionised Fe-line emission that is extremely
well-correlated with the continuum variations on timescales
$\sim 10$\,ks.

\section{The observations}

The datasets analysed were archived {\it XMM}
observations of duration
39\,ks from 2000 \citep{boller}
and 129\,ks from 2001 \citep{pr03}
together with the new observation
made during 2005 May 23 UT 19:21:51 - Jun
3 UT 21:27:10 over six {\it XMM} orbits, 
{
science observation IDs in the range 0304030[1-7]01.}
EPIC data
utilized the medium filter and pn ``PrimeSmallWindow'' mode.  
Our analysis here describes only the
pn data \citep{struder}, as these offer the best S/N ratio 
and were free of the photon pile-up effect.
Data were processed using SAS {\sc v6.5.0}. 
Instrument patterns 0--4 were selected. Source data
were extracted from a circular cell of radius $40''$, 
background data from a source-free region of the same pn
chip.  Periods when the full-band pn count rate exceeded 2 ct/s
in the background region or when the background exceeded 5\% of the
source count rate in the 5-10 keV band were excluded.
The screening yielded an effective 
exposure of 402\,ks over an observational baseline of $944$\,ks.
Mrk~766 gave a mean pn count rate 1.511$\pm0.002$ ct/s in the 
2-10\,keV band.  The mean screened background level was 1\%
of the mean source rate in this band. The 2005 mean 
2-10\,keV flux was
$F \sim 1.3 \times 10^{-11}{\rm erg\ cm^{-2}s^{-1}}$,
similar to the flux in 2000 May
($F \sim 1.2 \times 10^{-11}{\rm erg\ cm^{-2}s^{-1}}$) 
and lower than in 2001
($F \sim 2 \times 10^{-11}{\rm erg\ cm^{-2}s^{-1}}$) \citep{pr03}.

\begin{figure}
\centering{
  \resizebox{7.5cm}{!}{ 
  \rotatebox{0}{
  \includegraphics{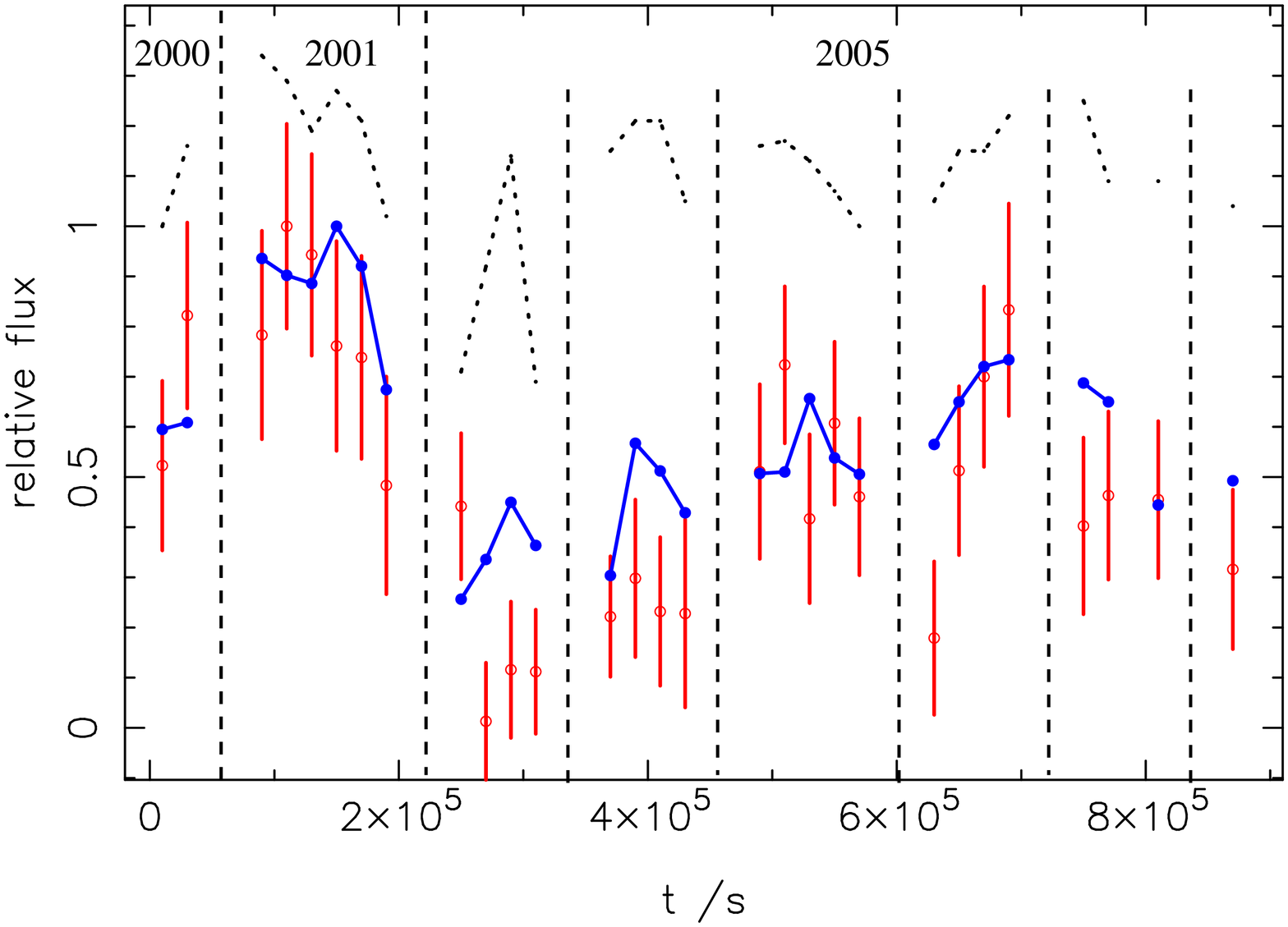}}}
}
  \caption{
Light-curves for the continuum
flux (blue curve, solid symbols) and emission-line flux (open, red
symbols, large error bars) in the high-ionisation band (observed
energy range 6.5-7.0\,keV). The data are shown in independent 20\,ks
intervals for the pn observations from 2000, 2001 and 2005.  
In this figure, an artificial time
interval of 40\,ks has been inserted between each dataset at the
times indicated by the vertical dashed lines.  The flux values of
each time series have been scaled to the maxima and
are given as "relative flux". Also shown is the variation of fitted
photon energy spectral index $\alpha$ (dotted line), which
has values in the range $0.7 < \alpha < 1.4$ as indicated on the y-axis.  
Typical uncertainty in $\alpha$ is $\pm 0.06$.
}
\label{fig1}
\end{figure}

\section{Analysis of the Fe-line variations}

To investigate the time variability of the Fe emission line we first
fit a continuum model to the pn data over the energy range
3-9.5\,keV, avoiding a strong soft-excess contribution at lower
energies. The energy range 6.0-7.0\,keV is excluded in order not to be
biased by the Fe line emission.  The model comprises a power-law of
varying amplitude and slope, absorbed by cold gas \citep{mm}
of variable column
density. The data are divided into independently-fitted 20\,ks
intervals and the background spectra in each interval 
subtracted.  Uncertainties on measured quantities include the
uncertainty in the background subtraction. The line emission in any
band is then estimated as the summed excess flux 
above the fitted continuum.  We consider 
two bands: a ``high-ionisation'' band of observed energy 6.5-7.0\,keV 
(rest energy 6.58-7.09\,keV) and a ``low-ionisation'' band of
observed energy 6.0-6.4\,keV (rest energy 6.08-6.48\,keV). The effects
of convolution with the instrumental response (FWHM $\sim 0.14$\,keV) and of
possible Doppler shifts (T06)
may produce some contamination between
high-ionisation and low-ionisation flux in each band: for brevity,
however, we use this nomenclature below.  

\begin{figure}
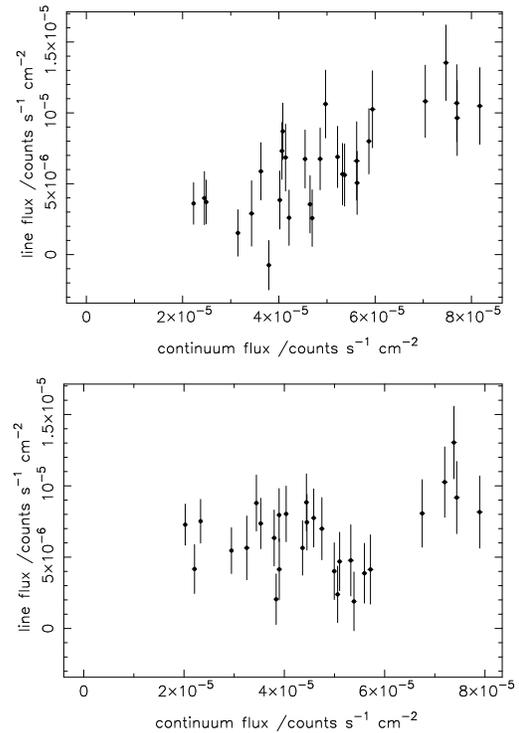

\centering{
  \begin{minipage}{6.5cm}{
  \resizebox{6.5cm}{!}{ 
  \rotatebox{-90}{
  \includegraphics{5276fig2a.ps}}}}
  \end{minipage}

  \vspace*{2mm}

  \begin{minipage}{6.5cm}{
  \resizebox{6.7cm}{!}{ 
  \rotatebox{-90}{
  \includegraphics{5276fig2b.ps}}}}
  \end{minipage}
}
  \caption{
{\em Top} The correlation
between continuum and emission-line flux in the high-ionisation band,
observed energy range 6.5-7.0\,keV, showing the same data points as in
Fig.\,\ref{fig1}.  The statistical uncertainty in continuum flux is much smaller
than that in the line flux. {\em Bottom} The lack of correlation between
continuum and emission-line flux in the low-ionisation band, observed
energy range 6.0-6.4 keV.
}
\label{fig2}
\end{figure}

Fig.\,\ref{fig1} shows the variation in
the emission-line flux in the high-ionisation band and the continuum,
quoted in the same band, from all epochs, 2000-2005.  For clarity, an
artificial interval of 40\,ks has been inserted between each dataset in
Fig.\,\ref{fig1}. There appears a strong correlation between the continuum and
high-ionisation emission-line variations. The correlation is
illustrated directly in Fig.\,\ref{fig2}, which shows a clear relationship
between continuum and high-ionisation emission-line flux but not
between continuum and low-ionisation emission-line flux.

\begin{figure}
\centering{
  \resizebox{6cm}{!}{ 
  \rotatebox{-90}{
  \includegraphics{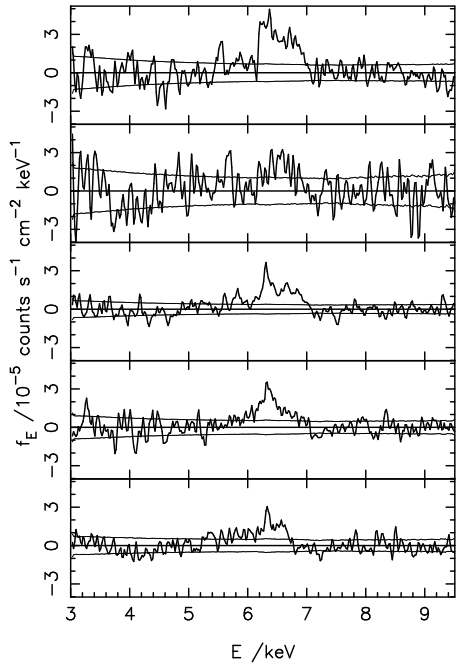}}}
}
  \caption{
Residual spectra above the fitted
continuum in the range $3-9.5$\,keV.  Data are divided 
into five flux states and the residuals within
each flux state co-added.  Flux states increase in flux
from the bottom panel to the top.  Spectra are sampled at
energy intervals of $0.04$\,keV and smoothed with a Gaussian of
FWHM $0.08$\,keV to reduce shot noise without 
significantly degrading resolution.
Lines above and below the x-axes show the rms values of the
random noise (this varies between panels because of the varying amount of
data incorporated into each flux state).
The y-axis scale is the same on each panel.
}
\label{fig3}
\end{figure}

The statistical significance of the
correlation may be estimated from the Spearman rank correlation test.
The rank correlation coefficient $r_s$ for the high-ionisation band has a
value $r_s=0.698$ with 29 data points for the 20\,ks sampling, $r_s=0.633$
with 57 data points for sampling at 10\,ks.  If the data points were
independent these values would have significance levels $p=1.3 \times 10^{-5}$ 
and $p=6.4 \times 10^{-8}$: 
however there is measurable autocovariance in both time
series which reduces the significance of the correlation between them
(although photon shot noise results in the emission-line time series
having an observed power spectrum that is almost flat).   
{
Hence we
estimate the significance level from $10^8$ pairs of time series
simulated from power spectra that match the data, and hence have the same 
autocovariance. The power spectra were estimated from a Monte-Carlo
likelihood fit to the data taking into account the window function and
shot noise. The simulations show that the probability
of obtaining the observed correlation by chance is
$p=2.3 \times 10^{-5}$ for 20\,ks or $p=10^{-7}$ for 10\,ks sampling.}
If we consider instead the low-ionisation band there is no correlation
(Fig.\,\ref{fig2}): for 20\,ks sampling, $r_s=0.158$ with significance level
$p=0.2$. There is weak evidence for the emission in this band being
variable, with significance level $p=0.05$ from a test of $\chi^2$, but any
variations do not appear strongly correlated with continuum
flux. Inspection of the residual spectra in Fig.\,\ref{fig3}, discussed below,
suggests that, although there is a narrow component of low-ionisation
Fe emission, there may also be a component of broadened, possibly
Doppler-shifted (T06), 
emission in this band during 2001.  

To test the origin of the correlation we inspect the spectra
of the residual flux above the fitted continuum.
In each individual
time interval the residual spectra are rather noisy, so to make a
sensitive investigation of the residuals we divide the data into five
"flux states", equally spaced in 2-10\,keV total flux between the
minimum and maximum values.  Each 20\,ks time interval is assigned to
one of these flux states and the residual spectra in each flux state
are co-added (Fig.\,\ref{fig3}).  Our aim is to investigate whether there is any
systematic error in the fitting procedure that may lead to a spurious
correlation, and dividing into flux states in this way is a good test
for this possible problem. It may be seen that the continuum has been
well-fitted over the 3-9.5\,keV energy range, and that there is indeed
a tendency for the emission-line flux in the high-ionisation band to
increase from the lowest to the highest flux states.  The
emission-line flux in the low-ionisation band does not show this
variation, as seen in Fig.\,\ref{fig2}. There are variations in the overall
emission-line profile, but we postpone more detailed modelling of the
spectrum to a later paper (Turner et al. in prep.).  

To further check that the ionised-band correlation is not an artefact of the 
continuum fitting procedure, we also consider the values of the fitted continuum model
parameters and the residuals about the model. 
Fig.\,\ref{fig1} also shows the
value of the fitted power-law energy index.  For most of the
observations the fitted value is consistent with being unchanging,
although a harder spectrum is found in the lowest flux
state at the start of the 2005 observations. 
Analysis of the continuum 
variations indicates that this arises from the presence of a hard
spectral component, visible in the lowest state in Fig.\,\ref{fig3}
that we interpret as reflection from
lower-ionisation material (Turner et al., in prep.).  
{
Whether the correlation
of Fig.\,\ref{fig2} passes through the origin depends on the modelling of
this component, but the strength of the correlation is unaffected by the
continuum model adopted.} For our
purposes here we wish to use a simple model solely to
measure the "local" continuum amplitude at $\sim 6.5$\,keV, not to establish
its physical origin, so have used a simple power-law parameterisation
of that. Repeating the correlation analysis omitting the entire orbit
comprising the lowest flux state changes the correlation coefficient
to $r_s=0.628$ with 25 data points, still significant at $p=2.8 \times 10^{-4}$.  

\begin{figure*}
\centering{
  \resizebox{14cm}{!}{ 
  \includegraphics{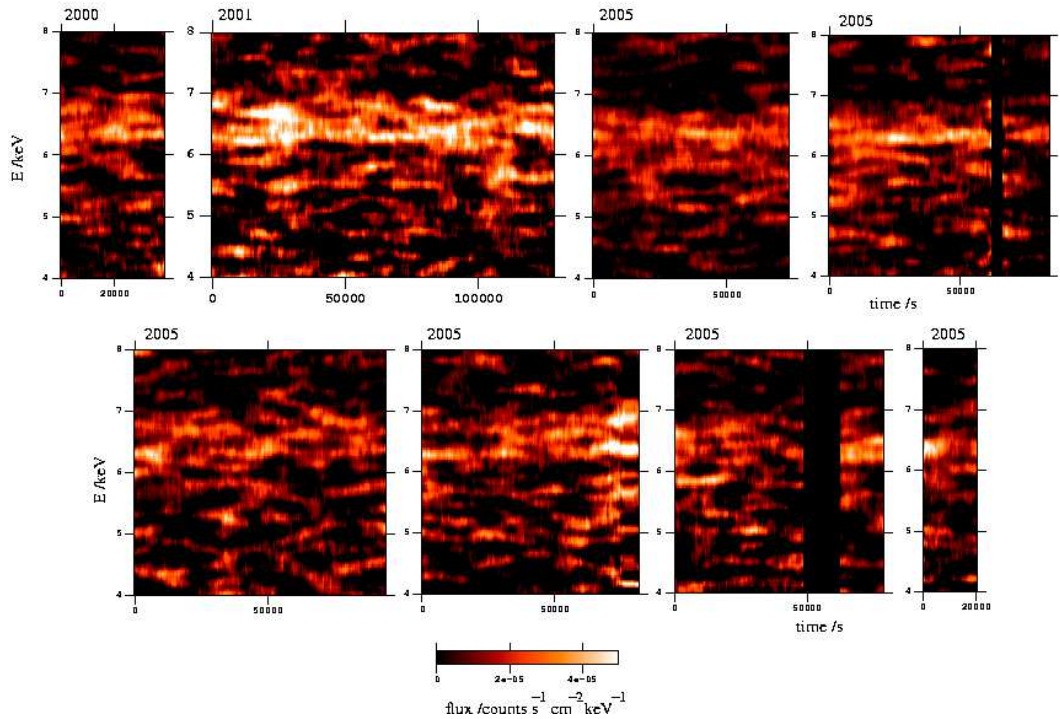}}
}
  \caption{
Colour-scale
representations of the varying source spectra are shown for each
observation, with the 2000
data in the upper left panel, then 2001 data, followed by six panels for the
2005 data. Photon energy is shown on the y-axis, time 
on the x-axis.  An absorbed power-law continuum model has been fitted
in 10\,ks time intervals and subtracted.
The scale of spectral flux density is indicated on the
colour bar. The time axis has been smoothed with a top-hat of width 10\,ks, 
the spectrum axis with a Gaussian of FWHM 0.14\,keV 
to match the instrumental spectral resolution (see also T06). 
Blank regions indicate where data with high background have
been removed. The strong ionised features characteristic of the 2001
data are largely absent in the 2005 data, although they 
appear towards the end of the fourth 2005 observation (sixth panel),
coinciding with the increase in flux seen in Fig.\,\ref{fig1}.
}
\label{fig4}
\end{figure*}

Having established a correlation between the high-ionisation line and
continuum variations, we attempt to constrain any possible time lag
between the two.  Such a lag would be a measure of the distance
between the line- and continuum-emitting regions, as used in
optical-ultraviolet ``reverberation mapping''. For a circularly
symmetric emission region (such as an ideal accretion disc) that is
centrally illuminated the time lag is a direct measure of the radius
at which the line radiation is produced.  Simulations of the data
yield weak evidence against the null hypothesis that there is a
correlation with zero lag, with significance $p=0.11$ and a best-fitting
time lag of 10\,ks, corresponding to a length scale of 20\,A.U., but
with a 68 percent confidence interval of $\pm 10$\,ks. A firmer upper limit
on the lag does not exist: at larger lag values the amplitude of the
window function is low (few data points overlap in the
cross-correlation) and the cross-correlation error becomes large.  If
confirmed by further data, the existence of a lag on ks timescales
would be consistent with the estimate for the emitting radius $r \sim 100 r_g$
found when analysing the energy shifts of the Fe line in the 2001
observation (T06).  

We have also searched the 2005 data for evidence of
significant energy shifts, however none are found.  Fig.\,\ref{fig4} shows
smoothed time-resolved spectra for the entire dataset: these spectra
look very different for the 2005 data, with no evidence for the
complex pattern of emission seen in the 2001 data (T06). 
The flux states
sampled in the 2005 observation are generally significantly lower than
in 2001, except for a period towards the end of the fourth orbit.
Fig.\,\ref{fig4} shows that prominent ionised-line features are only present in
the highest flux states, consistent with the correlation of Fig.\,\ref{fig2}.

\section{Conclusions}

Detection of a line-continuum correlation on timescales of
tens of ks makes it likely that the ionised Fe emission-line region is within
light-hours of the continuum source in Mrk~766. 
The accretion disc model provides
a good explanation of the general behaviour of active galaxies, and
here the correlated variability and the evidence for Doppler shifts of
the same gas (T06) lead us to conclude that this is indeed the expected
ionised reflection from an accretion disc. The variation timescale and
lack of a strongly gravitationally-redshifted asymmetric line profile
implies that the ionised emission is seen at tens to hundreds of
gravitational radii from the black hole. Future observations may allow
better determination of a time lag and hence direct measurement of the
X-ray-emitting accretion disc using reverberation mapping.

\label{lastpage}

\end{document}